\documentstyle[aps,pre,twocolumn]{revtex}  
  
\begin{document}
\draft

\title{Instance Space of the Number Partitioning Problem}
\author{ F.\ F.\ Ferreira and J.\ F.\ Fontanari}
\address{Instituto de F\'{\i}sica de S\~ao Carlos,
 Universidade de S\~ao Paulo\\
 Caixa Postal 369, 
 13560-970 S\~ao Carlos SP, Brazil 
}
%\date{}
\maketitle

\begin{abstract}
Within the replica framework we study analytically the instance
space of the number partitioning problem. This classic integer 
programming problem consists of partitioning a sequence of $N$ 
positive real numbers $\{ a_1,a_2,\ldots, a_N \}$ (the instance) 
into two sets such that the absolute value of the difference of the sums of 
$a_j$ over the two sets is minimized. We show that there is an
upper bound $\alpha_c N$ to the number of perfect partitions 
(i.e. partitions for which that difference is zero) and characterize 
the statistical properties of the instances for which those 
partitions exist. In particular, in the case that the two sets have 
the same cardinality (balanced partitions) we find $\alpha_c = 1/2$.
Moreover, we show that the disordered model resulting from  the instance
space approach can be viewed as a model of replicators where the random 
interactions are given by the Hebb rule. 
\end{abstract}


%-----------------------------------------------------------------------

\section{Introduction}\label{sec:level1}

Most statistical mechanics  analyses of 
combinatorial optimization problems have concentrated on the 
characterization of average properties of  minima of a given cost 
function \cite{GJ,MPV}. Usually, 
the cost function depends on  a large set of fixed parameters,
termed the instance of the optimization problem [e.g. the 
distances between cities in the celebrated travelling salesman 
problem (TSP)] which, in the framework of statistical mechanics,
are treated as quenched random variables 
distributed according to some known probability distribution.  
Furthermore, in order to consider the subspace of configurations 
with a given average cost, one defines a probability distribution
on the space of configurations (e.g. the $N!/2N$ different tours or 
ordering of the cities in the TSP), namely, the Gibbs
distribution with `temperature' $T=1/\beta$. 
The zero-temperature limit then singles out  the configurations 
that minimize the cost function (ground states). 
Clearly, in this formulation the 
configurations are treated as fast, annealed variables.

Instead, in this work  we explore the opposite viewpoint, namely,
given a set of configurations
we want to characterize the subspace of instances for which  
those configurations have a certain cost. This approach
may be viewed as a {\it best-case} analysis in the sense
that one searches for particularly easy instances that 
fit  the given solutions. 
The situation here is  similar to the physics approach to neural 
networks. In a first stage, attention was given to the neural dynamics
while the coupling strengths between neurons were kept fixed
according to some variant of the Hebb learning rule 
\cite{Hopfield,Amit}. (The neural dynamics 
itself can be viewed as a versatile  heuristic 
in which the optimization problem   
is embedded in the neural couplings \cite{Tank}.)
In the second stage which followed the seminal work of
Gardner \cite{Gardner}, the focus was on the characterization
of the couplings distribution that ensures the stability of
a given set of neural  states. Gardner's formulation allowed a
rich interchange of concepts and  methods between 
the statistical physics and the computational learning theory 
communities 
\cite{rev}.

The specific optimization problem we consider in this paper
is  the number partition problem (NPP) \cite{KKLO,JAMS}
 which has received 
considerable attention in the physics literature recently 
\cite{FFF,Mertens,Physa}. It 
is stated as follows.
Given a sequence of $N$ positive real numbers ${\bf a} =
\{ a_1,a_2,\ldots,a_N \}$
(the instance), the 
NPP consists of partitioning them into two disjoint
sets ${\cal {A}}_1$ and ${\cal {A}}_2$ such that the difference
\begin{equation}
\mid \sum_{a_j \in {\cal {A}}_1 } a_j 
- \sum_{a_j \in {\cal {A}}_2 } a_j \mid
\end{equation}
is minimized. Alternatively, we can search for the Ising spin
configurations ${\bf s} =
\{  s_1,\ldots,s_N  \} $ that minimize the cost function
\begin{equation}  \label{E_1}
E \left ( {\bf s} \right ) = ~ \mid \sum_{j=1}^N a_j s_j \mid,
\end{equation}
where $s_j = 1$ if $a_j \in {\cal {A}}_1 $ and $s_j = -1$ if 
$a_j \in {\cal {A}}_2 $.  Despite its simplicity, 
the NPP was shown to 
belong to the NP-complete class, which basically means that there 
is no known deterministic algorithm guaranteed to solve all instances 
of this problem within a polynomial time bound \cite{GJ}.

In the proposed framework,
we aim at characterizing the subspace of instances
$\{ {\bf a} \}$ for which  the fixed  set of partitions
$\{ {\bf s}^l \} ~l =1,...,P,$ are perfect, i.e., 
$E \left ( {\bf s}^l \right )=0 ~\forall l$.
To achieve this we define the  energy in the
instance space as 
\begin{equation}\label{e_0}
{\cal H} \left ( {\bf a} \right ) = 
\sum_{l=1}^{P} \left ( \frac{1}{\sqrt{N}}
\sum_{j=1}^{N}a_{j} s_{j}^{l} \right )^{2}
\end{equation}
so that the $P$ partitions are perfect only if ${\cal H} = 0$.
Henceforth we will assume that $P$ increases linearly with $N$,
i.e., $P = \alpha N$. Furthermore, we  assume that
the components $s_j^l$ are
statistically independent random variables drawn from the 
probability distribution
\begin{equation}\label{P_s}
{\cal P} \left ( s_{j}^{l} \right ) =
\frac{1}{2}\left( 1+\frac{m}{\sqrt{N}}\right) \delta
\left ( s_{j}^{l}-1 \right )+ \frac{1}{2}
\left( 1-\frac{m}{\sqrt{N}}\right) \delta
\left ( s_{j}^{l }+1 \right ) ,  
\end{equation}
where the weights of the Dirac delta functions are chosen
so that  $\langle s_{j}^{l} \rangle = m/\sqrt{N}$. The motivation
for this choice is twofold. First, the exhaustive search in
the Ising configuration space for $N \leq 26$
as well as the analytical solution of the linear relaxation
of the NPP indicate that the average 
difference between  the cardinalities of sets 
${\cal {A}}_1$ and $ {\cal {A}}_2$, 
\begin{equation}  \label{mt}
\hat{m} = \frac{1}{N} ~\sum_{j=1}^N s_j , 
\end{equation}
vanishes like
$1/ \sqrt{N}$ for large $N$ \cite{Physa}. 
Second, this scaling yields a non-trivial
thermodynamic limit, $N \rightarrow \infty$, for the average 
free-energy density associated to the  Hamiltonian (\ref{e_0}).

In this paper we will apply standard statistical mechanics techniques
to study analytically the ground states of the 
Hamiltonian  (\ref{e_0}). We concentrate our analysis  on the
zero-energy instances (i.e., instances for
which perfect partitions exist) only, since
the properties of the non-zero energy instances depend strongly on the 
rather arbitrary choice of the energy (\ref{e_0}).
Moreover,  perfect partitions are important from a
practical viewpoint as they may have code-breaking implications
\cite{Shamir} and so it may be of interest to
estimate the maximum number of perfect partitions that can be
encoded in an arbitrary instance, as well as to characterize
those instances that maximize the number of coded perfect 
partitions.

The rest of this paper is organized in the following way. 
In Sec.\  \ref{sec:level2} we use the replica method to evaluate
the average free-energy density in the thermodynamic limit 
and  to derive the replica-symmetric order parameters that describe
the statistical properties of the instance space. In particular,
we  show that there is a critical value, 
$\alpha_c \left ( m \right ) N$, which limits the number of perfect 
partitions. Also in that section, we study the stability
of the replica-symmetric solution against replica symmetry breaking 
and show that
the zero-energy instances can reliably be  described  by the
replica-symmetric order parameters.
In Sec.\  \ref{sec:level3} we calculate
the probability  density for a given entry, say $a_k$, 
to have value $a$. This is achieved by integrating the joint
probability distribution (the Gibbs distribution) over all 
entries except $a_k$.
Finally, in Sec.\ \ref{sec:level4} we present some concluding remarks.
In particular we show that the disordered model considered here is
formally identical to a model of replicators with the random interactions
given by the Hebb rule.

%-----------------------------------------------------------
\section{Replica approach}\label{sec:level2}
%-----------------------------------------------------------

Following the standard prescription of performing quenched averages
on extensive quantities only \cite{MPV}, we define the average
free-energy density $f$ as
\begin{equation}\label{f0}
- \beta f = \lim_{N \rightarrow \infty} \frac{1}{N} \left \langle 
\ln Z \right \rangle
\end{equation}
where 
\begin{equation}\label{Z0}
Z = \int_0^\infty \prod _j da_j \delta \left ( R - \frac{1}{N} \sum_j a_j
\right ) \mbox{e}^{- \beta {\cal H} \left ( {\bf a} \right ) }
\end{equation}
is the partition function and $\beta = 1/T$ is 
the inverse temperature. Taking the limit $T \rightarrow 0$ in 
Eq.\ (\ref{Z0}) ensures that only the instances that minimize 
${\cal H} \left ( {\bf a} \right )$
will contribute to $Z$.
Here $ \langle \ldots \rangle $ stands for the average over
the partitions ${\bf s}^l ~(l=1, \ldots, P )$. The constraint
on the mean of the instance vector is needed in order to exclude
the trivial solution ${\bf a} = 0$. Fortunately, the arbitrary parameter $R$
does not play any relevant role in the theory, giving only the scale of the 
order parameters of the model.

As usual, the quenched average in Eq.\ (\ref{f0}) is evaluated through
the replica method: using the identity
\begin{equation}
\left \langle  \ln Z \right \rangle = \lim_{n \rightarrow 0} \frac{1}{n} \ln
\left \langle  Z^n \right \rangle
\end{equation}
we first evaluate $ \left \langle  Z^n \right \rangle$ for {\it integer} $n$
and then analytically continue to $n=0$. Using standard techniques
\cite{GD} we obtain, in the thermodynamic limit
\begin{eqnarray}\label{f1}
-\beta f & = &  \lim_{n \rightarrow 0} \frac{1}{n} \mbox{extr}   \{
\frac{R^2}{2} \sum_\rho Q_\rho \hat{Q}_\rho - R^2
\sum_{\rho <\delta } q_{\rho \delta } \hat{q}_{\rho \delta} \nonumber \\
& & \mbox{} 
+R\sum_\rho \hat{R}_\rho +  \alpha \ln G_1 \left ( q_{\rho \delta },
Q_\rho \right ) \nonumber\\ 
& & \mbox{} 
+ \ln G_2 \left ( \hat{q}_{\rho \delta }, \hat{R}_\rho, \hat{Q}_\rho \right ) \} 
 \end{eqnarray}
where

\begin{eqnarray}\label{g1}
G_{1} & = & \int_{-\infty }^{\infty }
\prod_\rho \frac{d\tilde{x}_\rho}{\sqrt{2\pi} }
\exp \left [ -\frac{1}{2} 
\sum_\rho \left ( 1 + 2 \beta R^2 Q_\rho \right ) \tilde{x}_\rho^2 
 \right. \nonumber \\
& & 
\left. -2 \beta R^2 \sum_{\rho <\delta }\tilde{x}_\rho \tilde{x}_\delta 
q_{\rho \delta}+  ~ i m \sqrt{2 \beta R^2} \sum_\rho \tilde{x}_\rho \right ] 
\end{eqnarray}
and
\begin{eqnarray}\label{g2}
G_{2}=\int_{0}^{\infty } \prod_{\rho }da_{\rho }
\exp \left ( -\frac{1}{2} 
\sum_\rho \hat{Q}_\rho a_\rho^2 +\sum_{\rho<\delta }
\hat{q}_{\rho \delta }a_\rho a_\delta 
\right. \nonumber \\ 
\left.
-\sum_\rho \hat{R}_\rho a_\rho  \right ) .  
\end{eqnarray}

The extremum in Eq.\ (\ref{f1}) is taken over all order parameters  
$\left ( \hat{q}_{\rho \delta }, \hat{R}_\rho , \hat{Q}_\rho, 
q_{\rho \delta}, Q_\rho \right )$. The physical order parameters
\begin{equation}\label{q}
q_{\rho \delta} =  \left \langle \frac{1}{N R^2} \sum_{i=1}^N 
\left \langle a_i^\rho  \right \rangle_T  \left \langle a_i^\delta \right \rangle_T
\right \rangle  ~~~~~ \rho < \delta,
\end{equation}
and
\begin{equation}\label{r}
Q_{\rho} = \left \langle \frac{1}{N R^2} \sum_{i=1}^N 
\left \langle \left ( a_i^\rho \right )^2 \right \rangle_T \right \rangle
\end{equation}
measure the overlap between a pair of different equilibrium instances 
${\bf a}^\rho$ and ${\bf a}^\delta$, and the overlap of an 
equilibrium
instance ${\bf a}^\rho$ with itself, respectively. Here, 
$\langle \ldots \rangle_T$ stands for a thermal average.

\subsection{Replica-symmetric solution}

To proceed further we make the replica symmetric ansatz, i.e., we
assume that the values of the order parameters are independent of
their replica indices
\begin{equation}
\begin{tabular}{cccl} 
$q_{\rho \delta} = q $ & and & $\hat{q}_{\rho \delta} = \hat{q} $ 
& $\forall \rho < \delta$ \\
$Q_\rho = Q $ & and & $\hat{Q}_\rho = \hat{Q} $ 
& $\forall \rho $ \\
$\hat{R}_\rho = \hat{R} $ &  & 
& $\forall \rho $ . 
\end{tabular}
\end{equation}
Evaluation of Eqns.\ (\ref{g1}) and (\ref{g2}) with this ansatz is
straightforward. In order to write
the replica symmetric average free-energy density
it is convenient to introduce the new variables
\begin{equation}
\eta = R^2 \left (  \hat{Q} + \hat{q} \right ), ~~~~   
\tau = \frac{\hat{R}}{\sqrt{2 \left ( \hat{Q} + \hat{q} \right )}}, ~~~~
\theta = \frac{\hat{q}}{2 \left ( \hat{Q} + \hat{q} \right )} ,
\end{equation}
and rescale the temperature $ \beta'=\beta R^2$
so that  
\begin{eqnarray}\label{frs}
-\frac{\beta'}{R^2} f_{rs} & = &  
\frac{1}{2} \ln \left ( \frac{\pi R^2}{2} \right )
+ \frac{1}{2} \eta \left [ Q - 2 \theta \left ( Q -  q \right ) \right]
 +   \tau  \sqrt{ 2 \eta}
 \nonumber\\ & &
 - \frac{\alpha}{2} \ln \left [ 1 + 2 \beta' 
\left ( Q - q \right ) \right ] 
- \beta' \alpha 
\frac{m^2  + q }{1 + 2 \beta' \left ( Q - q \right )} \nonumber \\ & &
 - 
\frac{1}{2} \ln \eta
 +
\int_{-\infty}^\infty Dz \ln 
\left ( \mbox{e}^{\Xi_z^2} \mbox{erfc} \Xi_z \right )
\end{eqnarray}
where
\begin{equation}\label{Xi}
\Xi_z = \tau + z \theta^{1/2},
\end{equation}
and
\begin{equation}
Dz = \frac{dz}{\sqrt{2 \pi}} \mbox{e}^{-z^2/2}
\end{equation}
is the Gaussian measure. Thus it is clear from Eq.\ (\ref{frs}) that
the parameter $R$ yields the scales of the temperature and free-energy, not
affecting in any significant way the physical, replica-symmetric order 
parameters
\begin{equation}\label{qdef}
q = \left \langle \frac{1}{N R^2} 
\sum_{i=1}^N \left \langle  a_i \right \rangle_T^2 
\right \rangle  ,
\end{equation}
and
\begin{equation}\label{Qdef}
Q = \left \langle \frac{1}{N R^2} 
\sum_{i=1}^N \left \langle  a_i^2 \right \rangle_T 
\right \rangle  .
\end{equation}
The replica-symmetric
average energy density
$\epsilon_{rs} = \partial \left ( \beta f_{rs} \right )/ \partial \beta$
is given by
\begin{equation}\label{ers}
\epsilon_{rs} /\alpha R^2=  
\frac{q + m^2}{\left [ 1 + 2 \beta' \left ( Q - q \right )
\right ] ^2}
+  
\frac{Q-q}{1 + 2 \beta' \left ( Q - q \right )}
\end{equation}
which vanishes in the limit $\beta' \rightarrow \infty $ provided
that $q < Q$. 

\vspace{1.0 cm}
\begin{figure}[hbtp]
\vspace{7.cm}
\includegraphics{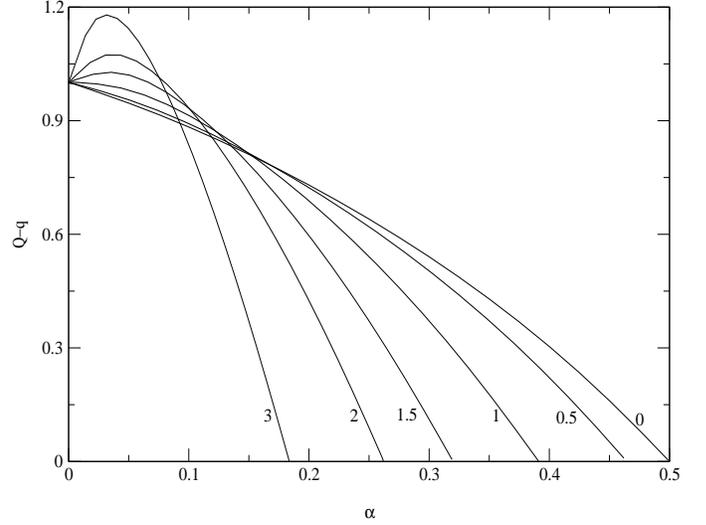}
\caption{\small{Average variance of the zero-energy instance entries
$Q-q$ as a function of $\alpha$ for $m=0$, $0.5$, $1$, $1.5$, $2$,
and $3$. The value of $\alpha$ at which the variance vanishes($\alpha_c$)
gives an upper bound to the number of perfect partitions.}}
\label{Fig.1}
\end{figure}

\begin{figure}[bhtp]
\vspace{7.cm}
\includegraphics{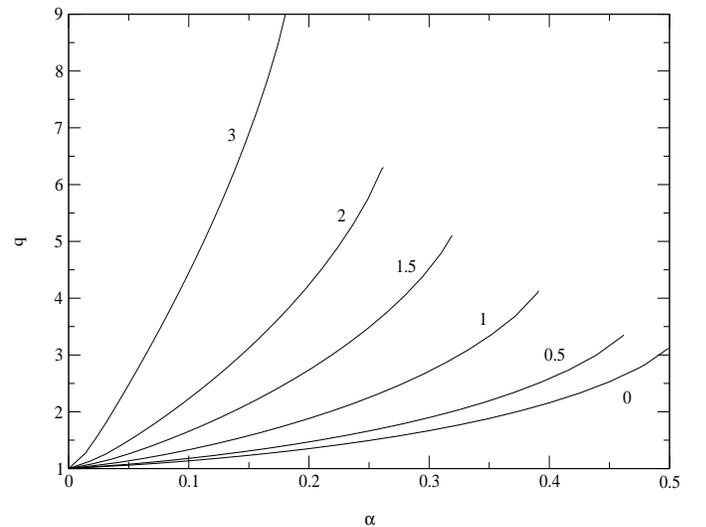}
\vspace{0.5cm}
\caption{\small{Average overlap between two different zero-energy
instances $q$ as
a function of $\alpha$ for $m=0$, $0.5$, $1$, $1.5$, $2$,
and $3$. The curves end at $\alpha = \alpha_c$.}}
\label{Fig.2}
\end{figure}
\vspace{1. cm}

As justified in Sec.\ \ref{sec:level1} we will focus
on this limit only. After some algebra, the saddle-point equations in
this limit are written as
\begin{equation}\label{s1}
\theta = \frac{q + m^2}{2 \left ( Q - q \right )} ~,
\end{equation}
\begin{equation}\label{s2}
\eta = \frac{\alpha}{ Q - q } ~,
\end{equation}
\begin{equation}\label{s3}
\tau = \sqrt{\frac{Q - q}{2\alpha}} \left ( 1 - \alpha + 
\frac{\alpha m^2}{Q-q} \right ) ~,
\end{equation}
\begin{equation}\label{s4}
 \sqrt{2 \eta} = - 2 \tau  + \frac{2}{\sqrt{\pi}}
\int_{-\infty}^\infty Dz ~\frac{\exp \left (-\Xi_z^2 \right )}
{ \mbox{erfc} \Xi_z } ~,
\end{equation}
\begin{equation}\label{s5}
\eta \left ( Q - q \right ) = 1  - \frac{1}{\sqrt{\pi \theta}}
\int_{-\infty}^\infty Dz ~z~\frac{\exp \left (-\Xi_z^2 \right )}
{ \mbox{erfc} \Xi_z } ~,
\end{equation}
with $\Xi_z$ given by Eq.\ (\ref{Xi}). 
In general these equations can be
solved numerically only. In Figs.~\ref{Fig.1}  and \ref{Fig.2} 
we present the dependence
of $Q- q$ and $q$, respectively, on $\alpha$ for different 
values of $m$.
For $\alpha = 0$ we find $Q - q = q =1$,  $\theta = (1 + m^2)/2$, and $\eta = 0$,
while $\tau$  diverges like $1/\sqrt{2 \alpha}$. 
According to the physical meaning of the order parameters given in 
Eqns.\ (\ref{qdef}) and (\ref{Qdef}), the difference $Q-q$ measures
the average variance of the zero-energy instance entries: 
the larger this difference, the larger the dispersion of the
instance entries. Interestingly, for fixed $m > 0$ this variance 
reaches its maximum for $\alpha > 0$. The divergence of the 
order parameters $Q$ and $q$ (and of their difference, as well),
for $m \rightarrow \infty$ and $\alpha \neq 0$ is expected, since
in order that an extremely unbalanced partition  become
a perfect partition there must exist some very large entries to 
compensate for the much larger number of entries in one of the sets.
Moreover, we observe from Fig.~\ref{Fig.1} that for fixed $m$ there is
a value of $\alpha = \alpha_c$ at which the overlap between two
zero-energy instances $q$ equals its maximal value $Q$.  
This results signals the shrinking of the
zero-energy instance subspace  to instances differing from a 
microscopic number
of entries $a_j$ only. Besides, it gives the limit of existence
of the solutions with zero-energy: for $\alpha > \alpha_c$ there
are no zero-energy instances. Taking the limit $q \rightarrow Q$
in the saddle-point equations (\ref{s1} - \ref{s5}) yields
\begin{equation}\label{a_c}
\alpha_c = 1 - \int_{- \Delta}^\infty Dz
\end{equation}
where
\begin{equation}\label{Rc}
\Delta = \left ( \frac{ \alpha_c m^4 }{Q_c + m^2} 
\right )^{\frac{1}{2}}
\end{equation}
is the solution of
\begin{equation}\label{D}
\frac{\Delta}{m^2} =   \int_{- \Delta}^\infty Dz \left ( z + \Delta
\right )  - \Delta . 
\end{equation}
Here $Q_c$ stands for the order parameter $Q$ evaluated at $\alpha_c$.
For $m=0$ we can solve these equations analytically:
we find that $\Delta$ vanishes like $m^2/\sqrt{2\pi}$ and so 
$\alpha_c = 1/2$ and $Q_c = \pi$. 

\begin{figure}[htpb]
\vspace{3.cm}
\includegraphics{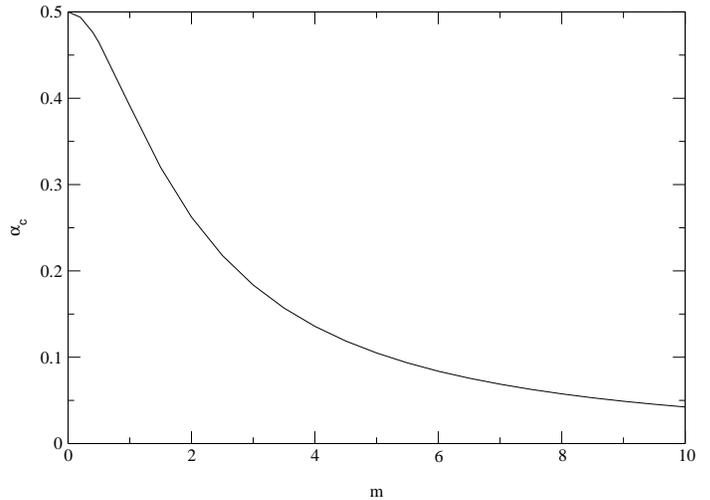}
\vspace{4.5 cm} 
\caption{\small{Instance independent upper bound to the number of perfect 
partitions $\alpha_c$ as a function of the parameter $m$ which measures
the unbalance of the partitions. For balanced partitions ($m=0$) we find
$\alpha_c = 1/2$.}}
\label{Fig.3}
\end{figure}

%\space{-2cm}
\begin{figure}[htpb]
\vspace{3.cm}
\includegraphics{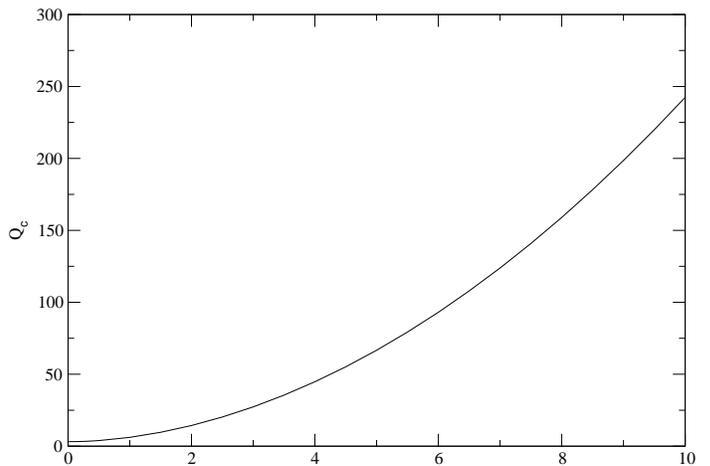} 
\vspace{4.0cm}
\caption{\small{Average  overlap of a zero-energy instance with itself
calculated at $\alpha_c$ as a function of $m$. 
For balanced partitions ($m=0$)  we find $Q_c = \pi$.}}
\label{Fig.4}
\end{figure}

In Figs.~\ref{Fig.3}  and \ref{Fig.4}
we show $\alpha_c$ and $Q_c$, respectively, as functions of $m$.
The dependence of $\alpha_c$ on $m$ indicates that
instances for which there are
an extensive number of unbalanced perfect partitions become
very rare with increasing $m$. In particular, there are no
zero-energy instances for partitions with average cardinalities 
difference [see Eq.\ (\ref{mt})] of order of $1$.

\subsection{Stability analysis}

The condition for the local stability of the replica-symmetric 
saddle-point  is given by \cite{GD}
\begin{equation}\label{stab0}
\alpha \gamma_1 \gamma_2 \leq 1
\end{equation}
where $\gamma_1$ and $\gamma_2$ are the transverse eigenvalues
\cite{AT} of
the matrices of second derivatives of $G_1$ and $G_2$ with respect
to $q_{\rho \delta}$ and $\hat{q}_{\rho \delta}$, respectively,
evaluated at the replica-symmetric saddle-point. After some algebra
we find that condition (\ref{stab0}) reduces to
\begin{equation}\label{stab1}
\alpha \, \left [ \eta \left ( Q - q \right ) \right ]^{-2} 
\int_{-\infty}^\infty Dz \left ( {\overline{a^2}} - 
{\overline{a}}^2 \right )^{2} \leq 1
\end{equation}
where
\begin{equation}
{\overline{a^n}} = 
\frac{\int_0^\infty da \; a^n \exp \left ( -\frac{1}{2} a^2 - 
a \;\sqrt{2} \; \Xi_z 
\right )}
{\int_0^\infty da  \exp \left ( -\frac{1}{2} a^2 - a \; \sqrt{2}
\; \Xi_z \right )} .
\end{equation}
Taking the limit $q \rightarrow Q$ we can easily show that 
\begin{equation}
 \eta \left ( Q - q \right) \rightarrow   \alpha_c 
\end{equation}
and 
\begin{equation}
\int_{-\infty}^\infty Dz \left ( {\overline{a^2}} - 
{\overline{a}}^2 \right )^{2} \rightarrow   \alpha_c ,
\end{equation}
with $\alpha_c$ given by Eq.\ (\ref{a_c}),  so that
the left hand side of Eq.\ (\ref{stab1}) equals
$1$ at $\alpha = \alpha_c$. Moreover,
we have verified numerically that this stability condition is 
always satisfied for $\alpha < \alpha_c $.

%-----------------------------------------------------------
\section{Probability distribution of entries}
\label{sec:level3}
%-----------------------------------------------------------

\begin{figure}
\vspace{6.cm}
\includegraphics{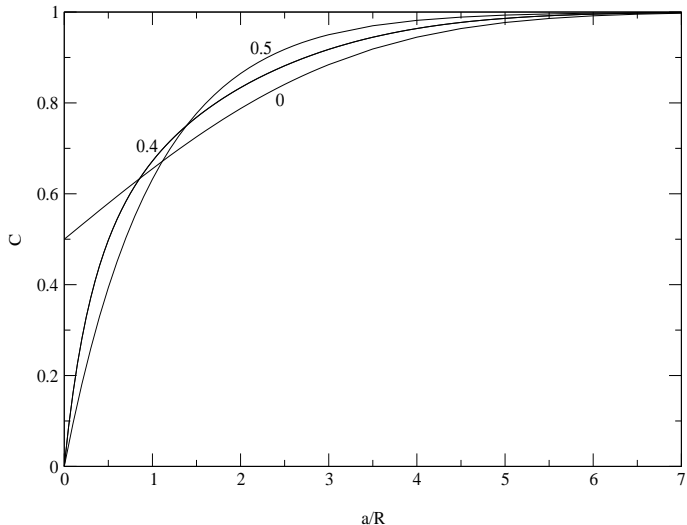} 
\vspace{2cm}
\caption{\small{Cumulative distribution function of the zero-energy
instance entries
for $m=0$ and  $\alpha = 0$, $0.4$, and $0.5$. Note that $C(0) \neq 0$
only at $\alpha = \alpha_c = 0.5$.}}
\label{Fig.5}
\end{figure}

The traditional probabilistic approach to study optimization 
problems introduces a probability distribution over the space of
instances. The main objection to this approach is that one rarely
knows what probability distribution is realistic. In the NPP,
for instance, it is usually assumed that the entries $a_k$ 
are statistically
independent random variables distributed uniformly in the unit
interval \cite{KKLO,JAMS,FFF}. In this section we calculate
analytically the distribution of probability that a certain entry,
say $a_k$, of a zero-energy instance assumes
the value $a$, defined by

\begin{eqnarray}\label{prob}
{\mathcal{P} }_k \left ( a \right ) &  = & \lim_{\beta \rightarrow \infty}
\left \langle ~ \left \langle \delta \left ( a_k - a \right )
\right \rangle_T ~ \right \rangle 
=   \lim_{\beta \rightarrow \infty} \left \langle ~
\frac{1}{Z} \int_0^\infty \prod _j da_j 
\right. \nonumber \\
& & 
\left. 
\delta 
\left ( R - \frac{1}{N} \sum_j a_j \right ) \  
\delta \left ( a_k - a \right )
\mbox{e}^{- \beta {\cal H} \left ( {\bf a} \right ) } \right \rangle
\end{eqnarray}

where $Z$ and  ${\cal H} $ are given by Eqns.\ (\ref{Z0}) and (\ref{e_0}),
respectively. As all entries are equivalent we
can write ${\mathcal{P} }_k \left ( a \right ) = {\mathcal{P} }\left ( a \right )
\forall k$. Hence to evaluate Eq.\ (\ref{prob}) we introduce the auxiliary energy 
\begin{equation}
{\cal H}_{aux} \left ( {\bf a} \right )  =  {\cal H} \left ( {\bf a} \right )
+ h \sum_k \delta \left ( a_k - a \right )  ,
\end{equation}
so that
\begin{equation}\label{der}
{\mathcal{P} }\left ( a \right ) = - \lim_{\beta \rightarrow \infty}  ~ 
\frac{1}{N \beta} \left. \frac{ \partial \langle \ln
Z_{aux} \rangle }{\partial h} \right |_{h=0}
\end{equation}
where $Z_{aux}$ is the partition function (\ref{Z0}) with ${\cal H}$ 
replaced by ${\cal H}_{aux}$. Of course, we note that the entries
$\left ( a_1, \ldots, a_N \right )$ are not statistically
independent and their joint probability distribution is simply the 
Gibbs  probability distribution
\begin{equation}\label{Gibbs}
{\mathcal W} \left ( {\bf a } \right ) = \frac{1}{Z} \exp \left [ - \beta
{\cal H} \left ( {\bf a} \right ) \right ] .
\end{equation}
As expected,  Eq.\ (\ref{prob})  is recovered by
integrating this joint distribution over $a_j $ for all $j \neq k$
and then setting $a_k = a$.  Using Eq.\ (\ref{der}) the 
calculations needed to evaluate
${\mathcal P} \left ( a \right )$ become analogous to those used 
in the evaluation of the free-energy density (\ref{f1}). Within the 
replica-symmetric framework and in the limit
$\beta \rightarrow \infty$ with $q < Q$ the final
result  is
\begin{eqnarray}\label{prob2}
{\mathcal{P} }\left ( a \right ) =\sqrt{ \frac{2 \eta}{\pi R^2}}~
\int_{-\infty}^\infty Dz ~\frac{1}{\mbox{erfc} \, \Xi_z} ~
\exp \left [ -\frac{ \eta}{2 R^2}~a^2 
\right.\nonumber \\
 \left. -a \Xi_z \, \left ( \frac{2 \eta}{R^2} \right )^{1/2} - \Xi_z^2
  \right ] ~~~~~~ a \geq 0, ~~~~~~~~~~~~~
\end{eqnarray}
which for $\alpha = 0$ reduces to
\begin{equation}
{\mathcal{P} }\left ( a \right ) = \frac{1}{R} \,
\exp \left ( -\frac{a}{R} \right ) ~~ a \geq 0 .
\end{equation}
To handle a possible singularity in the limit $q \rightarrow Q$ it
is more convenient to consider instead  the cumulative distribution 
function defined by
\begin{eqnarray}\label{cum}
{\mathcal{C} }\left ( a \right ) & = &
\int_0^{a} da' \, {\mathcal{P} }\left ( a' \right ) \nonumber \\
& = & 1 - \int_{-\infty}^\infty Dz ~
\frac{\mbox{erfc}\left [\Xi_z + a \left (\eta/2R^2 \right )^{1/2} \right ]}
{\mbox{erfc} \, \Xi_z} .
\end{eqnarray}
Taking the limit $q \rightarrow Q$ yields
\begin{equation}\label{cum_c}
{\mathcal C }_c \left ( a \right ) = 1 - \frac{1}{2} \mbox{erfc}
\left [ \frac{\Delta}{\sqrt{2}} \left ( 1  + \frac{a}{R m^2}
\right )  \right ]
\end{equation}
where $\alpha_c$ and $\Delta$ are given by Eqns.\ (\ref{a_c}) and 
(\ref{D}), respectively. The interesting feature of this distribution
is that ${\mathcal C }_c \left ( 0 \right )$ is non-zero indicating
thus that the probability distribution (\ref{prob2}) evaluated at 
$\alpha = \alpha_c$ has a delta peak in $a = 0$. Explicitly,
\begin{equation}
{\mathcal P }_c \left ( a \right ) =  {\mathcal C }_c \left ( 0 \right )
\, \delta \left ( a \right ) + \frac{d {\mathcal C }_c}{da}
   ~~~a \geq 0,
\end{equation}
which for $m=0$ reduces to
\begin{equation}
{\mathcal P }_c \left ( a \right ) =  \frac{1}{2}
\, \delta \left ( a \right ) + \frac{1}{\sqrt{4 \pi R^2}}
\exp \left ( -\frac{a^2}{4R^2} \right ) ~~~a \geq 0.
\end{equation}
In Fig.~\ref{Fig.5}  we show the  cumulative distribution function $C(a)$ 
for $m=0$ and several values of $\alpha$. We note that $C(0) \neq 0$
only at $\alpha = \alpha_c$. 

%-----------------------------------------------------------
\section{Conclusion}\label{sec:level4}
%-----------------------------------------------------------

Since the instance entries are continuous variables 
$a_i \in \left [ 0,\infty \right ) ~\forall i$ we can resort to a simple 
gradient descent algorithm to find the minima of the energy 
$ {\mathcal H} \left ( {\mathbf a} \right )$, defined by Eq.\ (\ref{e_0}), 
which have a given mean. More pointedly, it can be easily
verified that the dynamics
\begin{equation}\label{rep_dyn}
\frac{d a_k}{dt} = - a_k \left [ \frac{1}{N}\sum_{l=1}^P s_k^l \sum_{i=1}^N
s_i^l a_ i  - \frac{1}{NR} {\mathcal H} \left ( {\mathbf a} \right ) \right ]
~ \forall k
\end{equation}
minimizes  $ {\mathcal H} \left ( {\mathbf a} \right )$ while
the mean $\sum_i a_i$ is a constant of motion. Interestingly, Eq.\ ({\ref{rep_dyn})
is readily recognized as a particular realization of the classical replicator equation  
which has been used to describe
the evolution of self-reproducing entities (replicators) in a variety of
fields, such as game theory, prebiotic evolution and sociobiology,
to name only a few \cite{Schuster}. In fact,  $a_i$ can be viewed as the
concentration of species $i$ whose fitness ${\mathcal H}_i$ is the
derivative ${\mathcal H}_i = \partial {\mathcal H}/\partial a_i$ of a
fitness functional ${\mathcal H}\left ( {\mathbf a} \right )$. 
The (infinite) population of replicators is composed of $N$
different species which evolve under the constraint of constant total concentration.   
The disordered model considered here is a variant of the  model of 
replicators with random interactions studied by Diederich and Opper \cite{Opper}
(see also \cite{Biscari}) in which the fitness functional is given by
\begin{equation}\label{HDO}
{\mathcal H}_{DO} \left ( {\mathbf a} \right ) = \sum_{ij} J_{ij} a_i a_ j
\end{equation}
where the couplings $ J_{ij} =  J_{ji} ~\left ( i \neq j \right ) $ are
independent,
identically distributed Gaussian random variables with mean zero and variance $1/N$,
while the self-interactions are non-random, species independent control 
parameters of the model, i.e., $J_{ii} = u ~ \forall i$.  
Clearly, the energy given by Eq.\ (\ref{e_0}) can be rewritten in the form of 
Eq.\ (\ref{HDO}) with the  couplings given by the Hebb rule 
$J_{ij} = \frac{1}{N} \sum_l^P  s_i^l s_j^l $ \cite{Hopfield}.
We note that in our model the self-interaction  is $J_{ii} = \alpha ~\forall i$,
while the  mean  and  variance of the off diagonal couplings are 
$\alpha m^2/N \rightarrow 0$
and $\alpha \left ( 1 - m^4/N^2 \right )/N \rightarrow \alpha/N$, respectively.
 Moreover, as in the
Hopfield model \cite{Hopfield}, though the $s_i^l$ are independent random variables,
the couplings $J_{ij}$ are not. 
It is interesting thus to interpret our results in the light of the random
replicator model: for $\alpha < \alpha_c$ the global optimum of the fitness
functional, ${\mathcal H}\left ( {\mathbf a} \right ) = 0$, can be reached
with the coexistence of all species; for $\alpha = \alpha_c$ reaching that
optimum requires the extinction of a macroscopic number of species, as
signaled by the delta peaks in $a=0$;
and for $\alpha > \alpha_c$ the  interactions between species are such
that the optimum is never reached.

To conclude, we mention that while
the traditional approach of Computer Science to the
validation of combinatorial search algorithms
focuses almost exclusively on the instance space (e.g.
the {\it worst-case} analysis is basically a search for instances that 
give the poorest performance of the algorithm under 
study \cite{GJ}), the statistical mechanics approach
has concentrated mainly on the configuration space, with 
the instances being drawn from arbitrary probability distributions
\cite{MPV}. 
Building on the work of Gardner on neural networks 
\cite{Gardner}, we  illustrate in this paper the usefulness of
equilibrium statistical mechanics tools to investigate the
statistical properties of the instance space as well.
For the optimization problem  we have  considered,
namely, the number partitioning problem, we have searched the instance 
space for the best (easiest) instances  to show 
that there is a maximum number of uncorrelated perfect partitions, 
$\alpha_c \left ( m \right ) N$ (see Fig.~\ref{Fig.3}). In particular, 
for balanced partitions ($m=0$) we find $\alpha_c (0) = 1/2$.
Clearly, this result yields an upper bound to the number of
perfect partitions (ground-state degeneracy) that can be found 
for any arbitrary  instance.  
As in the neural networks case, the instance space analysis proposed 
in this paper can be extended to virtually all optimization 
problems.

\bigskip

We  thank Pablo Moscato
for illuminating conversations. The work of JFF was supported in part by 
Conselho Nacional
de Desenvolvimento Cient\'{\i}fico e Tecnol\'ogico (CNPq).
FFF is supported by FAPESP.

%----------------------------------------------------------------- 

%========================================================================

\end{document}